# The Tunable Hybrid Surface Phonon and Plasmon Polariton Modes in Boron Nitride Nanotube and Graphene Monolayer Heterostructures

Yu Sun[1,2,a)], Zheng Zheng[2)], Jiangtao Cheng[3)], Jiansheng Liu[2)]

1 School of Information Science and Technology, Beijing Forestry University, Beijing 100083, China

2 School of Electronic and Information Engineering, Beihang University, Beijing 100191, China

3 Department of Mechanical and Energy Engineering, University of North Texas, Denton, Texas 76207, USA

The hybrid modes incorporating surface phonon polariton (SPhP) modes in boron nitride nanotubes (BNNTs) and surface plasmon polariton (SPP) modes in graphene monolayers are theoretically studied. The combination of the 1D BNNTs and 2D graphene monolayer further improves the modal characteristics with electrical tunability. Superior to the graphene monolayers, the proposed heterostructures supports single mode transmission with lateral optical confinement. The modal characteristics can be shifted from SPP-like toward SPhP-like. Both the figure of merit and field enhancement of hybrid modes are improved over 3 times than those of BNNT SPhP modes, which may further enable sub-wavelength mid-infrared applications.

Mid-infrared (IR) has great potentials in chemical and bio-molecular sensing, free space communication and thermal imaging for both civil and military purposes[1]. However, most of these potential applications have been limited by the lack of mid-IR devices comparable to those in the near-IR regime[2]. Surface polaritons are electromagnetic waves coupled to material charge oscillations, including the surface plasmon polaritons (SPPs) and surface phonon polaritons (SPhPs)[3]. Both SPPs and SPhPs exhibit high field enhancement and deep sub-wavelength confinement, which enables the next generation of mid-IR optical devices[4].

The SPPs, collective oscillations of free electrons, are accessible in noble metals at visible and near-IR frequencies[5]. Recently, the SPPs have been extended to the mid-IR and THz spectral ranges by graphene[6], a two dimensional (2D) layer of carbon atoms packed in hexagonal lattice[7]. The SPhPs originate from the phonon resonances at the mid-IR wavelengths in polar dielectric crystals, such as SiC, $SiO_2$ and hexagonal boron nitride (hBN)[8,9]. The multiwall boron nitride nanotubes (BNNTs), a one-dimensional (1D) version of BN, supports SPhP modes with higher field confinement and enhancement than those at flat interfaces[10]. Both the effective index of 2D graphene SPP modes and 1D BNNT SPhP modes are capable of reaching ~70 in mid-IR frequencies[5,6].

Although the BNNTs provide a suitable nanoscale geometry to support single-mode transmission with lateral field confinement, the modal properties can only be statically controlled through the nanotube size or the supporting substrates[11].

a) Email: sunyv@bjfu.edu.cn

The graphene monolayer can be dynamically tuned by the applied electric field[5], but the 2D structure could not provide lateral field confinement. The graphene ribbon waveguide supports lateral optical confinement, yet only the extreme narrow ribbon can operate at single-mode region[5,12]. It is challenging to fabricate, integrate and excite such a narrow ribbon.

To circumvent the above problems, we present the tunable hybrid modes in heterostructures incorporating 1D BNNTs and underlying 2D graphene monolayers. The hybrid modes originate from the plasmon-phonon coupling[13]. The hybrid modes exhibit high field confinement and enhancement. Superior to the graphene monolayers, the proposed heterostructures supports single mode transmission with lateral optical confinement. By tuning the chemical potential of graphene, the characteristics of hybrid modes can be shifted from SPP-like toward SPhP-like. By optimizing the chemical potential of the underlying graphene, the hybrid modes are capable of improving both the figure of merit and field enhancement over 3 times than those of BNNT SPhP modes. The hybrid modes may further enable mid-infrared applications far below the diffraction limit.

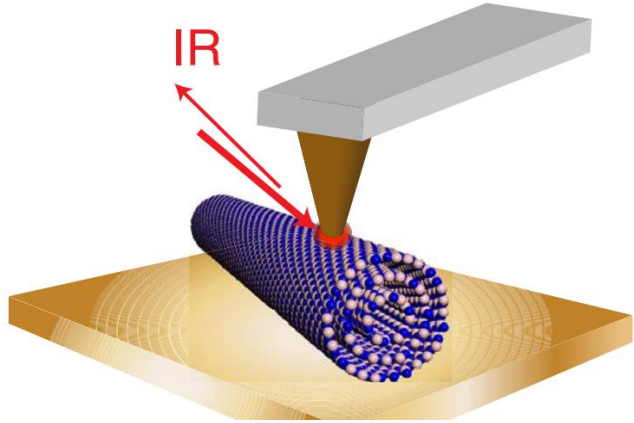

FIG. 1. Schematic of the proposed waveguide. (SCREENSHOT, TO BE UPDATE IN VER2)

The schematic of proposed heterostructures is shown in Fig. 1. The BNNT (25 nm diameter, 1/3 hollow)[13] is separated from an underlying graphene monolayer by a 5nm air ($n=1$) gap[14]. An oscillator model is used to describe the dielectric function of hBN[5]. To focus on the unique tunability of graphene, the geometric parameters are fixed in this letter. The graphene is uniformly biased and tuned by the applied electric field using metallic gates shown in Fig. 1. The substrate is KCl ($n=1.466$)[5], with air covering the rest regions. Graphene's complex conductivity is governed by the Kubo formula, which relates to the wavenumber $\upsilon$, charged particle scattering rate $\Gamma$, temperature $T$, and chemical potential $\mu_c$.[15] The value of $\Gamma = hq_eV_F/(2\mu E_F)$ is estimated from the measured impurity-limited DC graphene mobility $\mu \approx 10{,}000$ cm$^2$/(Vs),[16] where $E_F$ is the Fermi levels, $h$ is the Planck's constant, $q_e$ is the charge of an electron, $V_F$ is the Fermi velocity (~$10^8$cm/s in graphene).[17] The complex effective index $n_{eff}$-

$ik_{eff}$ and field profiles of the hybrid modes are investigated at $\upsilon$=1,420cm$^{-1}$, $T$=300K by the commercial finite-element method (FEM) solver COMSOL$^{TM}$. Convergence tests are done to ensure that the boundaries and meshing do not interfere with the solutions.

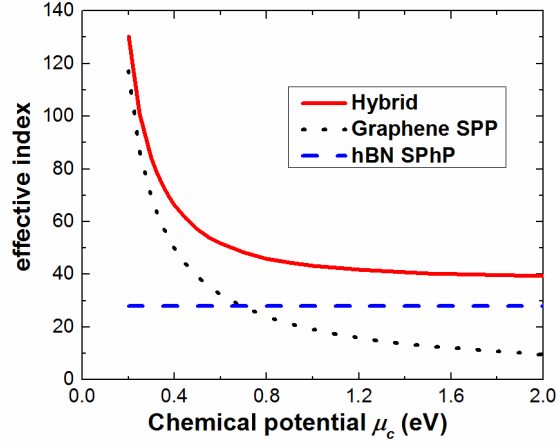

FIG. 2. Dependence of effective index on $\mu_c$ for the hybrid modes and un-hybrid SPP/SPhP modes.

The chemical potential of graphene $\mu_c$ can be tuned flexibly by external gates. The effective index $n_{eff}$ of the hybrid mode with different $\mu_c$ are shown in Fig. 2. For illustration, the effective index of un-hybrid SPP modes supported by individual 2D graphene (without the BBNT in Fig. 1) and un-hybrid SPhP modes supported by individual 1D BBNT (without the graphene in Fig. 1) are also depicted. As shown in Fig. 2, the hybrid modes evolve from SPP-like to SPhP-like asymptotically with increasing $\mu_c$. The BNNT operates in single mode condition, so only one SPhP mode can be excited and hybrids with graphene SPP mode. The single mode transmission is also confirmed by mode search using COMSOL$^{TM}$.

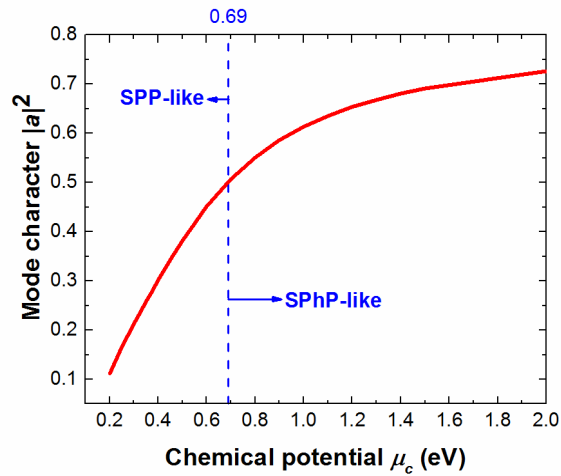

FIG. 3. Dependence of $|a|^2$ on $\mu_c$ for the hybrid mode as descripted by the coupled mode theory.

This hybridization can be modeled as the electromagnetically coupled oscillation between SPP and SPhP modes.[9] The polarization originates from lattice displacement in the BNNT exerts a force on the free carriers of graphene via near field interaction, and likewise the polarization originates from carrier displacement in the graphene exerts a force on the BNNT lattice. By tuning the chemical potential of graphene, the coupling can become constructive. Then the modal characteristics of the two constitutive polariton modes can be significantly enhanced and resulting the hybrid modes[18]. In order to gain a deeper understanding, the hybrid the modes are analyzed by coupled-mode theory[19]. The hybrid modes are modeled as a superposition of the graphene SPP modes and the BNNT SPhP modes: $n_{eff}(\mu_c)=an_{spp}(\mu_c)+bn_{sphp}$, where $a$ and $b$ are the amplitude of constituent SPP and SPhP modes respectively. The square norm of the SPP mode amplitude $|a|^2$ is a measure of hybridization:

$$|a|^2 = \left| \frac{n_{eff}(\mu_c) - n_{spp}(\mu_c)}{[n_{eff}(\mu_c) - n_{spp}(\mu_c)] + [n_{eff}(\mu_c) - n_{sphp}]} \right|. \quad (1)$$

In this respect, the modes are SPhP-like for $|a|^2>0.5$ and SPP-like otherwise. The dependence of $|a|^2$ on $\mu_c$ is shown in Fig. 3. The profile of $|a|^2$ confirms the transition between SPP-like and SPhP-like modes. The Figs. 4(a-d) depicts evolution of the electric field ($|E|$) profiles of hybrid modes with increasing $\mu_c$. As shown in Fig. 4, the concentrations of $|E|$ field shift from the underlying graphene toward the upper BNNT with increasing $\mu_c$. The $|E|$ profile of un-hybrid 1D BNNT SPhP modes are presented in Fig. 4(e) for comparison. The Fig. 3 also indicates the point of critical coupling: $\mu_c=0.69eV$. The $|E|$ field profiles of critical coupling are shown in Figs. 4(c). At the critical coupling point, the hybrid modes manifest equal SPP and SPhP characteristics ($|a|^2=|b|^2=0.5$), corresponding to $n_{spp}(\mu_c)=n_{sphp}=28.08$, indicating that two kinds of surface polaritons move in phase[20,21]. The modes are highly confined and enhanced in the nanoscale air gap between the BNNT and graphene monolayer.

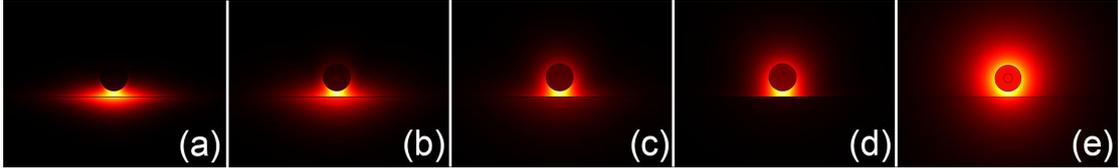

FIG. 4. $|E|$ profiles of hybrid modes with (a) $\mu_c=0.20eV$, (b) $\mu_c=0.35eV$, (c) $\mu_c=0.69eV$, (d) $\mu_c=2.00eV$; $|E|$ profile of un-hybrid 1D BNNT SPhP modes.

The figure of merit ($FoM=n_{eff}/2\pi k_{eff}$)[20] and field enhancement of the hybrid modes are shown in Fig. 5. The field enhancement is defined as the maximum of the electric field magnitude for a mode carrying 1μW power along the $x$ axis shown in Fig. 1[20]. By tuning the chemical potential of graphene, the figure of merit reaches the maximum value 2.37 when $\mu_c=0.35eV$. The corresponding field enhancement is up to $2.25\times10^7$V/m, which indicates that the field strength is dramatically enhanced as shown in Fig. 4(b). For comparison, the figure of merit and field enhancement of un-hybrid 1D BNNT SPhP mode is 0.69 and $7.18\times10^6$V/m, respectively. By the hybridization with the graphene SPP mode，both the figure of merit and field enhancement of BNNT SPhP mode are improved more than 3 times.

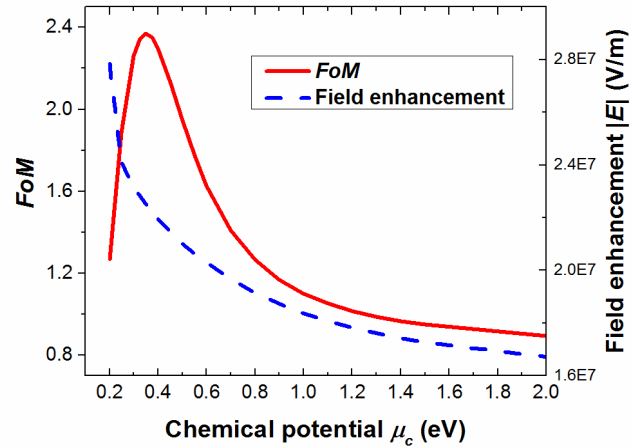

FIG. 5. Dependence of figure of merit (*FoM*) and field enhancement on $\mu_c$.

In this letter, we report the tunable hybrid surface phonon and plasmon polariton modes supported by heterostructures incorporating BNNTs on a graphene monolayer. The combination of the 1D BNNTs and 2D graphene monolayer provides an avenue to further improve modal characteristics with electrical tunability. Superior to the graphene monolayers, the proposed heterostructures supports single mode transmission with lateral optical confinement. By optimizing the chemical potential of the underlying graphene, the hybrid modes is capable of improving both the figure of merit and field enhancement over 3 times than those of BNNT SPhP mode. The hybrid modes may further enable mid-infrared applications in guiding, concentrating and harvesting optical energy far below the diffraction limit.

This work was supported by the Fundamental Research Funds for the Central Universities (BLX2014-26/TD2014-01), 973 Program (2012CB315601) and NSFC (61435002/61107057/61221061/51378156).